# Scanning-free three-dimensional fluorescent dipoles imaging by polarization self-interference digital holography (pSIDH)

Tianlong Man1*, Wenxue Zhang1, Lu Zhang1, Ran Zheng2, Hua Huang2, Xinhui Liu2, Hongqiang Zhou1, Zhe Wang3, Yuhong Wan1*

1School of Physics and Optoelectronic Engineering, Beijing University of Technology, Beijing, 100124, China

2Faculty of environment and life, Beijing University of Technology, Beijing, 100124, China

3 Consiglio Nazionale delle Ricerche, Istituto di Scienze Applicate e Sistemi Intelligenti, Italy

Correspondences: Tianlong Man (t.man@bjut.edu.cn), Yuhong Wan (yhongw@bjut.edu.cn)

**Abstract:** Polarization microscopy provides insights into the structure and orientational organization of biomolecules and their architectures in cells. The above key functional signatures, which are natively 3D, can be only detected in 2D for a single measurement in conventional polarization microscopy. It is so far a challenging task to capture simultaneously the 3D structure and molecular orientation in a single frame of far-field intensity distribution, within the timescale of rapid-happened spatial organization events of bio-complexes. We report an optical imaging method called pSIDH, to encode multidimensional sample information includes 3D structures and dipole orientations, in their far-field fluorescence-self-interference pattern. The computational reconstruction from the holographic extracted complex-valued light field provides optical-aberration-corrected 3D polarization images of the sample. In pSIDH microscope incorporating planar liquid crystal lens and high numerical aperture objective, we demonstrate scanning-free 3D volumetric polarization imaging of fluorescently-labelled sample, with simultaneously computational-improved system measuring accuracy on the 3D spatial and polarization dimensions. The pSIDH imaging on phalloidin-fluorophore labelling U2OS cells provides rapid tools of capturing simultaneous the 3D structural details and spatial-averaged molecular orientation distributions of biological complex architectures such as actin filaments.

**Keywords:** Digital holography, Multi-dimensional imaging, Fluorescence microscopy, Polarization microscopy, Computational imaging

**Introduction**

The precise way that proteins arrange in space and its spatial-temporal organizations play a vital role in biological functions of cells and tissues. Numerous cell mechanics driving essential biological processes, including the cell development and immune response, are powered by the dynamic remodeling of actin and/or septin filament[1–3]. Yet the direct protein organization measurement methods such as electron microscopy and X-ray diffraction techniques are however not applicable to live cell imaging. Alternatively, the precise structural organization information of the biomolecules can be obtained from the three-dimensional positions and orientations of the rigidly linked fluorophores. Since most fluorescent emitters exhibit a polarization dependent absorption and emission, the precise orientation information of the fluorophores can be in practical extracted from the polarization state of the excitation light and/or the emitted fluorescence[4]. Two strategies of fluorescence polarization microscopy (FPM) have been developed successfully: polarized single-molecule localization microscopy (SMLM) and microscopy with polarization projection techniques.

In polarized SMLM both the spatial position and molecular orientation information are encoded in the intensity distribution of the point spread function (PSF) of the system[5–10]. However, it is a challenge to further improve the imaging accuracy because of the intrinsic coupling of the spatial position and orientation, and the temporal resolution of the system is limited by the imaging mechanism of SMLM as well. In FPM based on polarization projection the signals from different polarization channels are used to reconstruct both the image and orientation distribution of the sample[11–14]. This approach has led to recent success in polarized structured illumination microscopy that can provide the fluorescent dipole information of cellular structures[14]. Nevertheless, rapid volumetric imaging on the dynamics of cellular structures still remains a challenge for current FPM due to the insufficient 3D imaging temporal resolution that originated from the axial-scanning-based image acquisition mechanism.

Fluorescence self-interference digital holography (SIDH) is a non-scanning 3D computational imaging technique[15]. By using a beam splitting phase mask generated with spatial light modulator (SLM), the complex wavefront of the light field that originated from each point source (fluorophore, for example) within a spatial incoherent or self-illuminate sample can be recorded in the hologram. 3D intensity distributions of the object can be then reconstructed computationally from only a single frame of the complex-valued hologram. In general, these holographic wavefront detection and reconstruction mechanism for incoherent imaging system provides unique advantages in 3D fluorescence imaging such as the inherent superior temporal resolution, sub-diffraction-limited lateral spatial resolution[15–18], and the computational optical aberrations compensation ability[19,20]. Combing the linear polarization splitting of SLM with wave plates, recent successful demonstration has been made in encoding the state of polarization of the object light field in the SIDH holograms[21].

All the above attempts have shown the potential of SIDH on the simultaneous reconstruction of the 3D spatial and polarization information of the sample. Nevertheless, the application of the SIDH in 3D polarization imaging of biomolecules has not been discussed yet. The limitation comes from the fact that the sophisticated GRIN[17] or custom-designed birefringent[18,22] beam splitting elements is indispensable for the high-resolution SIDH microscope but hard to be fabricated.

   Here, we report an optical imaging method called pSIDH, to encode multidimensional sample information includes 3D structures and dipole orientations in their far-field fluorescence-self-interference pattern. The 3D intensity and polarization orientation distributions of the sample, which are retrieved respectively from the amplitude and phase of the complex-valued holographic reconstructions, are demonstrated to be independent with each other and therefore enables high quality non-scanning polarization imaging of 3D samples. The validity of the imaging method was demonstrated in proof-of-principle experiments. A high numerical aperture (NA) pSIDH microscope incorporating commercially available planar liquid crystal lens (PLL) as polarization holographic splitting element was specifically designed, to provide the ability of scanning-free volumetric polarization imaging of the 3D distributed nanoparticles and immunofluorescent-labelled U2OS cells. The results demonstrated that the computational adaptive optics (CAO) optimization, which enabled by the wavefront recording and reconstruction ability of pSIDH, can concurrently improves the 3D spatial resolutions and signal-to-noise-ratio (SNR) of holographic intensity reconstructions as well as the polarization orientation detection accuracy of holographic phase reconstructions. We demonstrate the capabilities of our pSIDH microscope in rapid multidimensional imaging by simultaneously reconstructing the 3D structural and dipole orientation information of phalloidin-fluorophore labelling F-actin filaments forming stress fibers in the fixed U2OS cells. Furthermore, theoretical discussions are presented and showed that for the wobbling fluorophores the mean orientation can be uncoupled from their fluctuations by sequential capturing two holograms in the proposed method.

## Results

**Performance validation of pSIDH in 3D scanning-free polarization imaging.** Though scanning-free 3D imaging and even polarization imaging have been achieved in the SLM-based SIDH systems, and the using of SLMs gives flexibility to the SIDH systems while different type of beam splitting masks can be rapidly exchanged. This, however, comes at the cost of low photon efficiency and extra system complexity, therefore decreases the system performances and makes it even unpractical to apply the SIDH in one of its most intriguing applications, fluorescence microscopy. The instead adopting of high-efficiency beam-splitting elements elevate significantly the performances of SIDH and makes it possible to image the biological cells. Nevertheless, the custom-designed birefringent crystal or GRIN-based elements are however difficult to be fabricated and obtained, as well as bulk compensating optical elements are necessary to correct the large optical path difference. Recently progresses in real-time high-quality 3D imaging of macro-objects by SIDH have shown the potential of liquid crystal planar lens (or geometric phase lens) as a compact, high-efficiency beam-splitting elements. We noticed that, comparing with the combination of SLM and waveplates, the inherent circular (rather than linear) polarization splitting ability of PLL could make it proper as even superior polarization information encoder in SIDH system. Fig. 1**a** illustrates the basic concept on the 3D spatial coordinates and orientation encoding mechanism in our pSIDH method when imaging an individual fluorescent dipole. For an extended object assumed to be composed by numerous incoherent points, the amplitude and phase distribution of the holographic reconstruction can be used to retrieved the 3D structures and dipole orientation information, respectively. To demonstrate this, we carried out experiments on a proof-of-principle apparatus (Supplementary Fig. S**1**). Briefly, pSIDH system with PLL as beam-splitting element was built to encode simultaneously the 3D structural and polarization information of the object light in the holograms. For a linear polarized object that has flexible polarization orientations, the 3D coordinates of each point on it are encoded in the complex wavefront that been captured by the differential interference of the two beams, hence the object can be reconstructed in 3D by numerical back-propagating the holograms to different axial planes. Meanwhile, the polarization orientations are converted to phase difference between the RCP (right hand circular polarized)- and LCP (left hand circular polarized)-beams after the PLL and finally a linear term in the phase distribution of the reconstructed complex-valued images of the object (Materials and methods). This quantitative correspondence, which is of important, was first verified in the experiments. The combination of a back-illuminated resolution test target and a polarizer was used to simulate the polarized object with controllable orientations and imaged in the system. After a linear calibration procedure where the raw phase values were inverted and projected to the range of -$\pi/2$ to $\pi/2$ (Materials and methods), the phase of the numerical refocused reconstructions exhibit a linear dependence on the polarization orientation angles of the input light (Supplementary Fig. S2**a**), with a statistical accuracy (Standard deviation, std.) of about <0.6° (Supplementary Fig. S2**b**, Materials and methods). Therefore, the proposed pSIDH system present a scanning-free multidimensional imaging method, while from a single frame of complex-valued hologram the spatial structures of the sample are visualized from the intensity of the reconstructed images (Fig. 1**b-d**), and simultaneously high accuracy polarization orientations measurements by converting the phase values on each pixel in the reconstructed images to the orientation angles through look-up-table that obtained by a linear fitting on the data shown in Fig. S2**a** (Fig. 1**e**-**g**, Materials and methods).

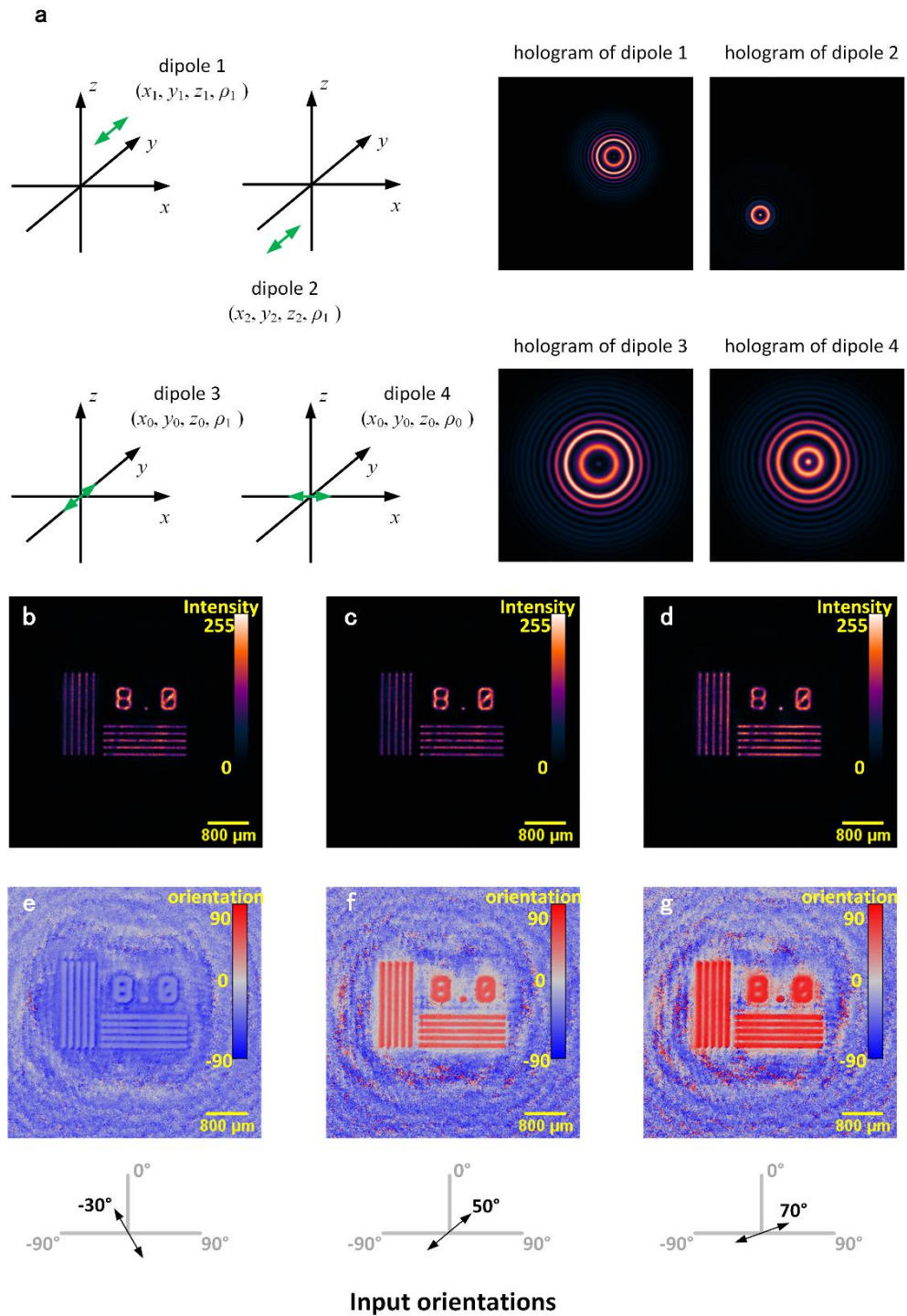

Fig. 1 Proof-of-principle validation of the pSIDH in simultaneously reconstruction of the intensity distribution and polarization orientation of the sample. **a** Schematic illustration of the 3D spatial coordinates and dipole orientation encoding mechanism in pSIDH. For an individual fluorescent dipole, the coordinates in $(x, y)$ plane are encoded in the central position of its hologram, while the coordinate along $z$ is encoded in the fringe density of the hologram. Meanwhile, the dipole orientation is encoded in the relative phase shifting of the hologram. **b-d** Reconstructed intensity and **e-g** orientation angle distribution of the sample (combination of resolution test target and polarizer) at different orientations of the polarizer.

Another important performance of pSIDH is that whether the above two independent physical parameters can be fully decoupled from each other, especially for the imaging of 3D object when the reconstruction is performed by numerically propagating the light field sequentially to different axial planes to obtain the 3D image stacks. We therefore imaged an object with fixed polarization orientation but flexible axial position in the sample space, and measured the dependence of both the reconstruction distance (specific distance used in the back propagation when the reconstructed object structures are numerically refocused, determined by using the auto-focusing algorithm, Materials and methods) and retrieved polarization orientations on the axial locations of the object. Experimental results (Fig. 2**a**) indicated that the reconstruction distances grow proportionally to the increases of the object axial location, which demonstrates the 3D imaging ability of the proposed method. Meanwhile, the retrieved orientations are proved to be insensitive to the axial locations, hence demonstrated that in pSIDH the object polarization orientations (linear phase term in the holograms, Materials and methods) can be independently decoupled from its 3D positions (quadratic phase term in the holograms, Materials and methods).

As the result, with the rapid pSIDH computational volumetric imaging the polarization orientation dynamics at different depths within the sample volume can be measured (Fig. 2**b**, **e**, **f**), simultaneously with the spatial structures (Fig. 2**c**, **d**). It should be emphasized again that in pSIDH only a single frame of complex-valued hologram of the sample is required to recover both the 3D spatial structures and polarization orientations.

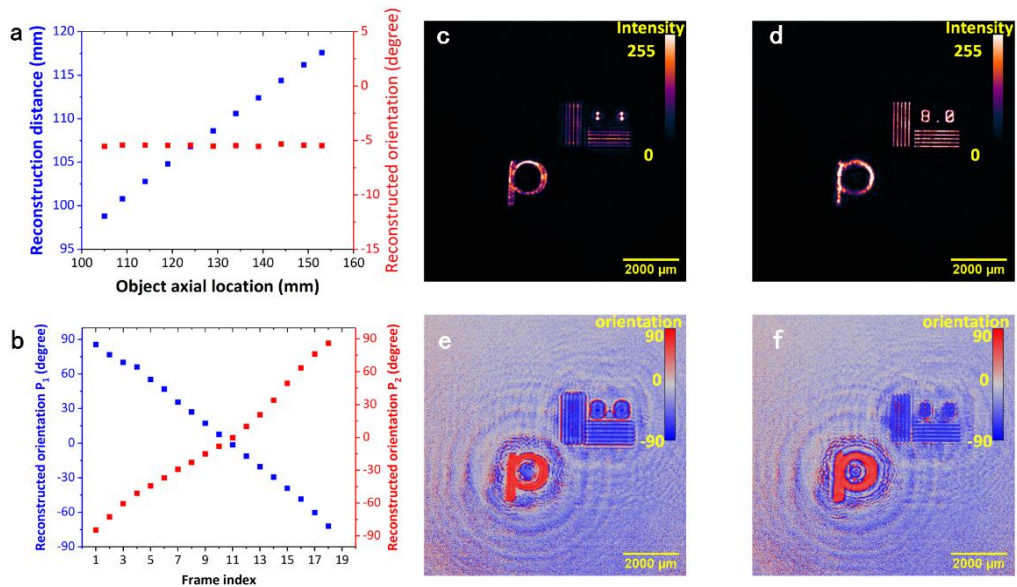

Fig. 2 Demonstrations on the 3D polarization orientation imaging by pSIDH. **a** The holographic reconstructed 3D positions and polarization orientations of the sample are proved to be independent. **b** Reconstructed polarization orientation dynamics at two different imaging depths within the sample volume. Numerical refocusing simultaneously obtained **c**, **d** intensity and **e**, **f** orientation (calibrated phase) images recover the spatial distributions of structures and polarization orientations at different imaging depths (holographic reconstruction distances of -97.6-mm for the letter "P" and -102.8-mm for the resolution test target, respectively) of the 3D sample. In the experiments two 2D polarized sample that located at different axial planes (100-mm and 114-mm to the objective lens respectively in the sample space), with their polarization orientations rotated along clockwise and counter-clockwise directions respectively during the measurements, are used to simulate a 3D

polarized dynamic sample. A pSIDH system with two independent sample channels was used and 18 frames of complex-valued holograms were captured sequentially (Materials and methods).

**pSIDH enables scanning-free, high-fidelity 3D computational adaptive optics (CAO) polarization fluorescence microscopy.** To explore the potential of pSIDH in high-resolution rapid 3D polarization orientation imaging of biomolecules, we designed pSIDH microscope with high NA objective lens (Fig. 3**a** and 3**d**, NA=1.3). In general, a home-build wide-field (WF) fluorescence microscope was modified, by adopting a commercially available high-throughput PLL (PBL25-532-F2500-SP, LBTEK) as SIDH holographic wave-dividing device (Materials and methods). The basic imaging performances of the system were quantified by imaging the resolution test target (1951 USAF hi-resolution fluorescence target, Materials and methods). By numerical back propagating the hologram (Fig. 3**b**, here the cropped amplitude distribution of the complex-valued hologram was shown, Materials and methods) to a proper distance, the high spatial resolution and SNR refocused image of the sample can be obtained with an overall magnification of 117× and pixel size (projected in the sample space) of 110.7-nm over an FOV of 113-μm × 113-μm (Fig. 3**c**, Materials and methods). The utilization of PLL (instead of SLM) improves the light energy efficiency and minimizes the insert optical aberrations, resulting in significant elevating on the imaging performance of SIDH especially here in a high NA microscope. Comparable performances were observed, with only minor degradations on the SNR, when the object was located below or above the optimal holographic recording plane (Supplementary Fig. S**3**, Materials and methods).

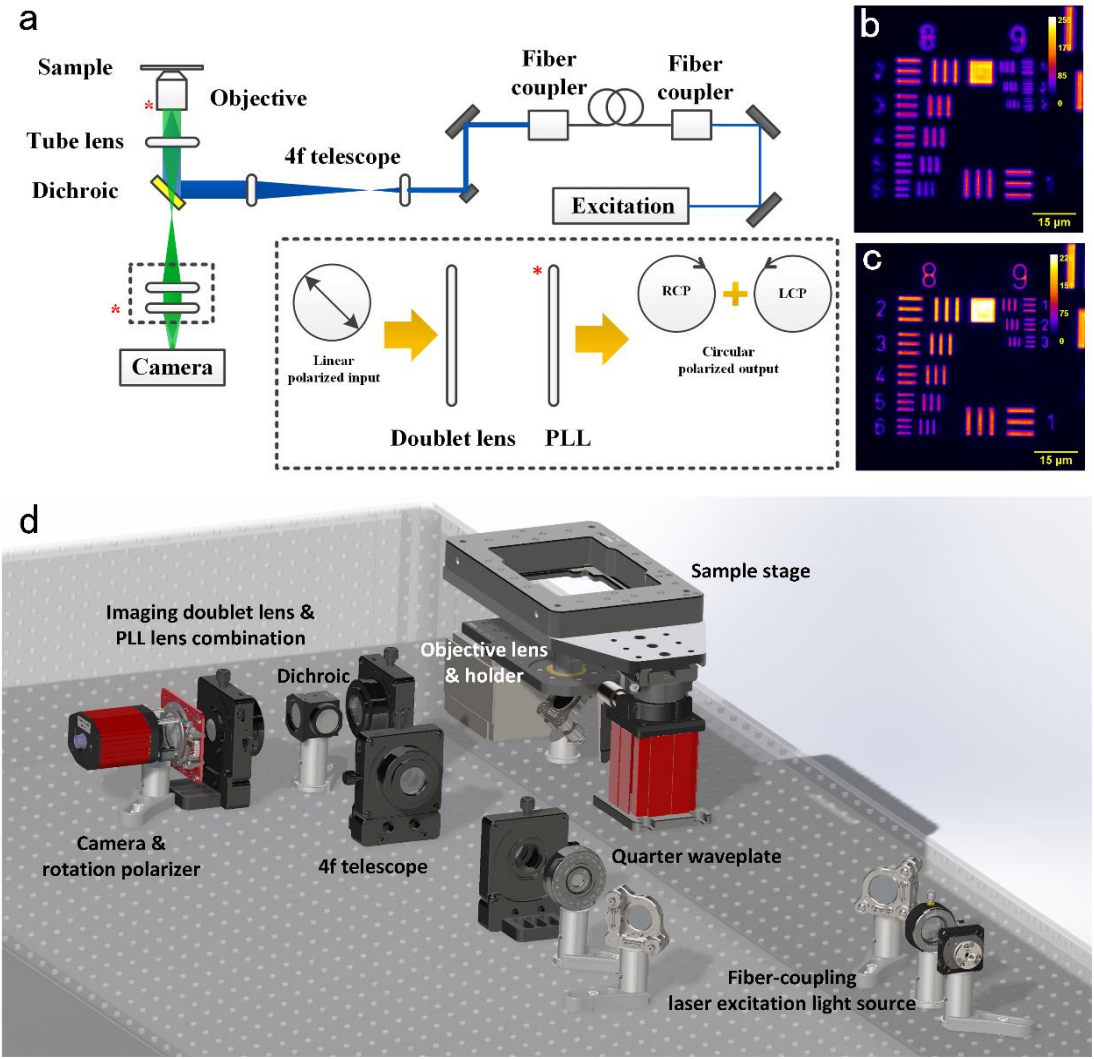

Fig. 3 Experimental results demonstrated the high spatial resolution, high SNR imaging ability of pSIDH microscope with high NA objective lens. **a** Schematic of pSIDH microscope. The red asterisk indicates the optical conjugation between the back pupil of the objective and the PLL (Materials and methods). **b** The captured hologram is back propagated to obtain the **c** numerically refocused reconstructed image of the sample. The images were cropped and the proper holographic reconstruction distance (here $z_r$ = -30-mm) was determined using auto-focusing algorithm (Materials and methods). **d** 3D representation of the pSIDH microscope in the lab.

To quantified the volumetric imaging ability of pSIDH microscope, we embedded 0.5-μm fluorescent particles in agarose (Materials and methods) and captured holograms. The phase-shifted holograms were processed to calibrate the transverse drifts and any variances in the total fluorescence intensities (Materials and methods). 3D WF image stack of the sample was recorded in the same system, for the calibration of the inherent anisotropic voxel of SIDH along the axial direction ($z$) in the image space (Materials and methods). From the holograms (Fig. 4**a**) the 3D distribution of the particles (Fig. 4**b**, **c**) was reconstructed, over an FOV of 113-μm × 113-μm × 100-μm and voxel size of 110.7-nm × 110.7-nm × 400-nm, within acquisition time of ~1.2-s. The 3D point-spread-functions (PSFs) in pSIDH exhibit an averaged axial full-width-half-maximum (FWHM) of ~4.4-μm, slightly wider than that of in the WF (~3.25-μm). Surprisingly, though the two modalities have the similar voxel sizes after calibration (Materials and methods), the pSIDH provides a sharper PSFs in lateral direction (FWHM of ~320-nm) than the WF (FWHM of ~497-nm, see Supplementary Fig. S**4** for quantitative comparisons). This can be explained by the super-resolution imaging ability of SIDH in the lateral direction. It should be noticed that the lateral resolution of WF was underestimated since the diameter of the fluorescent particles (500-nm) is larger than the theoretical FWHM of the PSFs (~240-nm). The degradations of axial resolutions (~610-nm theoretically) for both pSIDH and WF are mainly come from the inevitable spherical aberrations when using the oil immersion objective.

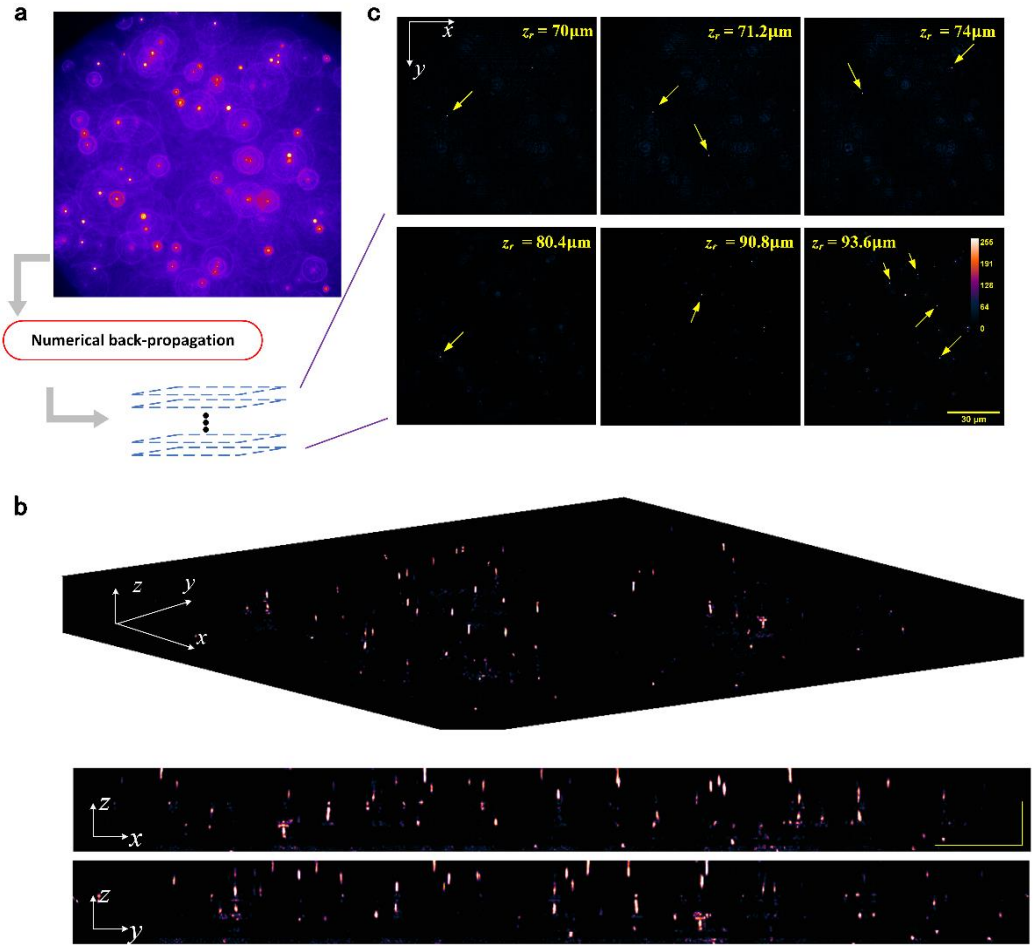

Fig. 4 Reconstruction of the fluorescent particles in 3D demonstrate the scanning-free volumetric imaging ability of pSIDH microscope. From the **a** captured hologram the **b** 3D distribution of the particles (cropped) is retrieved via a 3D numerical back-propagation. **c** At each depth slice refocused particles (indicated by the inserted arrow) are observed. Scale bar in **b**: 10-μm along $x$ and $y$, 5-μm along $z$.

In conventional microscope the compensating of optical aberrations is accomplished by using bulk AO unit including Shack-Hartmann sensor and SLM or deformable mirror, or in an iterative image optimization procedure where additional exposures are required. In our pSIDH microscope, on the other hand, this can be carried out via computational adaptive optics (CAO) without any additional hardware in the system or extra photon dose on the samples. After a CAO optimization that specifically addressed on spherical aberrations (Materials and methods), pSIDH exhibit a minor improvement on the lateral resolution (with FWHM of PSFs in *xy*: ~360-nm and ~350-nm before and after CAO, respectively), while the axial resolution is enhanced substantially (with FWHM of PSFs in *xz*: ~3-µm and ~2.1-µm before and after CAO, respectively), for volumetric reconstructions of fluorescent particles that located at different 3D coordinates in sample space (Fig. 5**a**, **b**). To our knowledge this is the first demonstration of SIDH scanning-free 3D volumetric imaging that has a large 3D FOVs in a high-NA system.

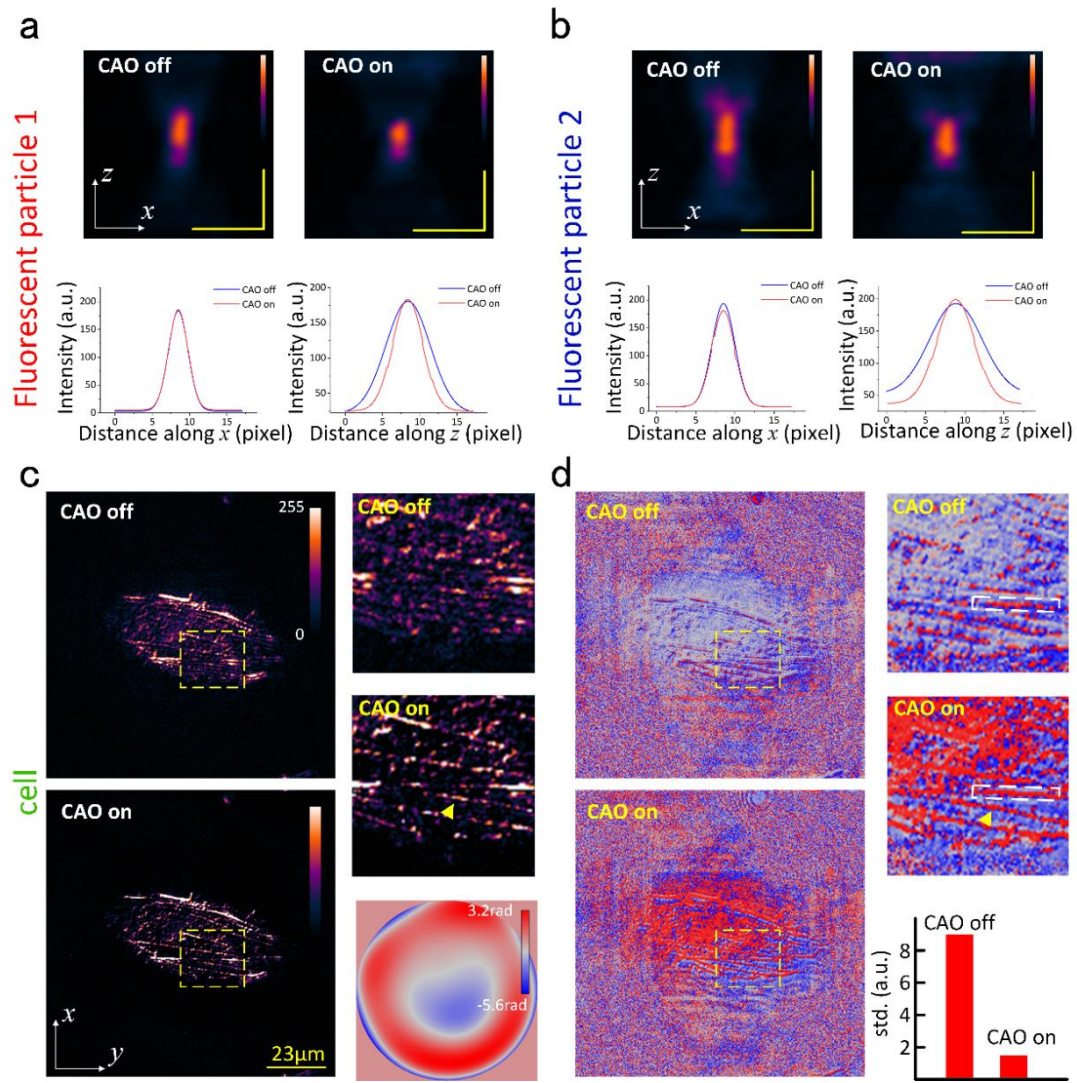

Fig. 5 Computational adaptive optics (CAO) pSIDH improves simultaneously the 3D spatial resolutions, SNR and measuring accuracy on the spatial-averaged molecular dipole orientations. CAO optimization (specifically applied on spherical aberrations) on the 3D pSIDH reconstruction enhances the optical resolutions along all the three spatial dimensions, in **a**, **b** volumetric images of fluorescent particles that located at different 3D coordinates in sample space (FWHM of PSFs before CAO: 342.1-nm laterally and 2.8-μm axially in **a**, 380.8-nm laterally and 3.15-μm axially in **b**; After CAO: 332.1-nm laterally and 1.92-μm axially in **a**, 379.7-nm laterally and 2.29-μm axially in **b**). After applying CAO optimization (first $20^{th}$ Zernike polynomials excepts tip, tilt and defocus were used) on the pSIDH complex-valued reconstruction of fluorescently labelled actin filaments of U2OS cells, **c** the calibrated intensity reconstructions uncover more structural details, bottom-right inserted: phase mask that been applied on the virtual pupil plane to compensate the aberration in CAO, Materials and methods. Meanwhile **d** the calibrated phase reconstructions improve the accuracy in the measuring of the spatial-averaged molecular orientations, bottom-right inserted: std. of the spatial-averaged phase values within the areas indicated by the white-colored dashed boxes (9.19 before and 1.49 after applying the CAO, lower is better, Materials and methods). The emerged individual actin filaments that brought back by the CAO optimization were marked with yellow arrowheads. Scale bar in **a**, **b**: 1-μm along $x$ and $y$, 3.2-μm along $z$.

In practice, more complex imaging conditions and hence more complicated optical aberrations challenge the high-quality microscopic visualization of biological samples (e.g., actin filaments within cells). Fortunately, the missed structural information in the conventional microscope due to the image degradation that caused by the optical aberrations, on the other hand, can be retrieved in our CAO pSIDH method. After applying CAO optimizations on the first 20$^{th}$ Zernike polynomials (excepts tip, tilt and defocus, Materials and methods) both the intensity and phase distributions of the pSIDH reconstructed complex-valued fluorescent light fields exhibit superior imaging performances. The improvements on 3D spatial resolution and SNR were indicated by the resolving of individual actin filaments (marked with yellow arrowheads in Fig. 5**c**) in the CAO calibrated pSIDH intensity reconstruction results. Meanwhile, CAO calibrated values along the actin filaments in the pSIDH phase reconstructions, which encode the fluorescent polarization and further the spatial-averaged molecular dipole orientations, were demonstrated to have a lower std. and therefore better measuring accuracy (Fig. 5**d**). It should be noted that for the results in Fig. 5 a polarizer was inserted before the PLL in the experiments. As the result, the samples emit a linear-polarized fluorescence and the dipole orientations of all the fluorescent molecules on the actin filaments are enforced to be parallel with the orientation of the polarizer. Therefore, in principle the pSIDH reconstructed phase values on the actin filaments should present a low-level of spatial deviations. Consequently, besides the spatial structural information, CAO pSIDH simultaneously improves the measuring accuracy on the spatial-averaged molecular dipole orientations.

**Simultaneously reconstruction of 3D structure and dipole orientations of biological samples using pSIDH.** The phase of the holographic reconstructions was calibrated in pSIDH microscope by imaging a polarized fluorescence resolution test target with controlled polarization orientation (Supplementary Fig. S**5**, Materials and methods). Similar with it is in the proof-of-principle system, the phases of the refocused reconstructions exhibit a linearly dependence on the input orientation angles. When combing with the pSIDH 3D imaging of the fluorescence intensity, these enable the simultaneously holographic reconstruction of both the structural and spatial-averaged dipole orientation information of the biomolecules inside the cells, after labelling the protein of interest with rigidly attached fluorescent dipoles emitter. The potential of our pSIDH microscope as a powerful tool in multidimensional bio-imaging was demonstrated by measuring the organization of actin filaments in cells using a modified system (Materials and methods). The actin filaments of U2OS cells were labelled with Alexa Fluor 488-phalloidin and then imaged under the pSIDH microscope. The captured phase-shifted holograms were processed to compensate the stage- and sample-introduced-drifts (Materials and methods), and then used to calculate the raw complex-valued hologram. The spatial organizations of the actin filaments can be reconstructed in 3D (Top row of Fig. 6: max intensity projection in $xy$ plane, middle and bottom rows of Fig. 6 present the computational refocused and calibrated intensity of the reconstructions at different axial planes, Materials and methods), together with the spatial-averaged molecular dipole orientations (computational refocused and calibrated phase of the reconstructions, Materials and methods), from a single frame of complex-valued hologram of the sample, by using the 3D holographic reconstruction and CAO optimization algorithm as described in details above.

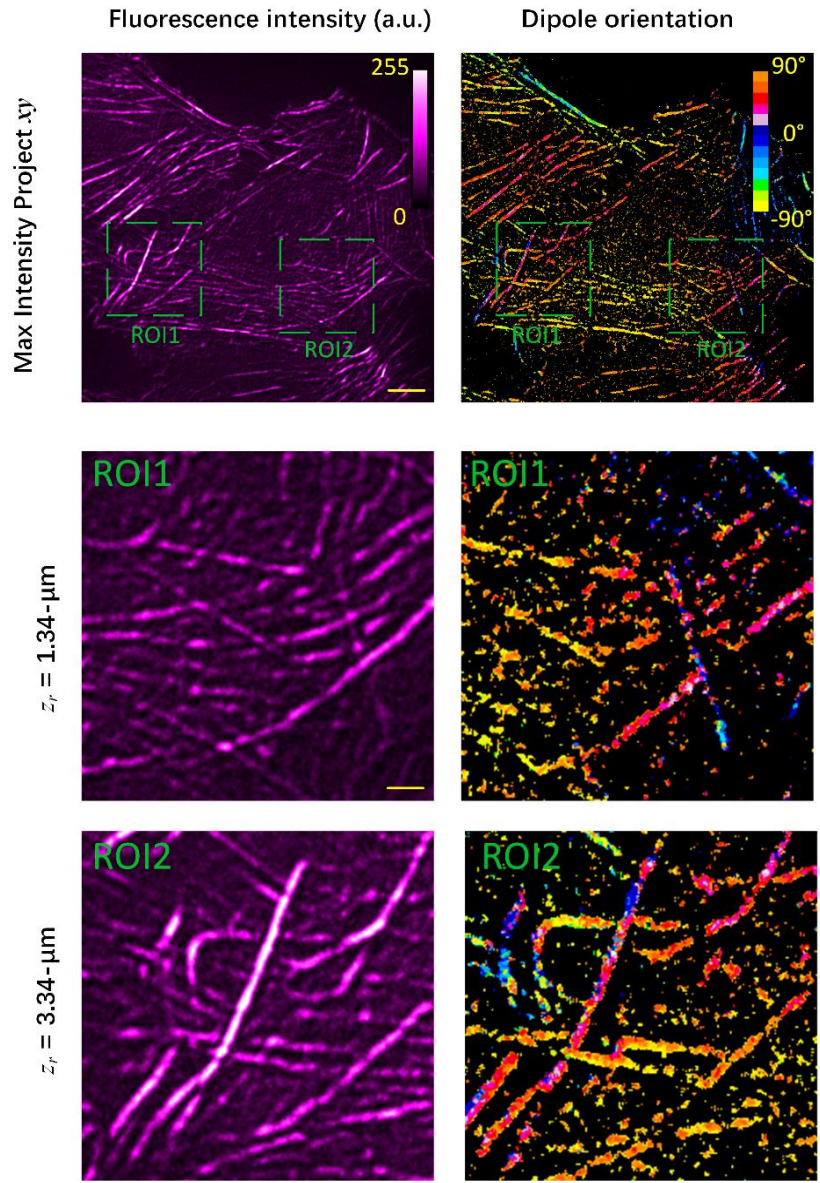

Fig. 6 pSIDH enables scanning-free, high-fidelity 3D CAO dipole orientation imaging of Alexa Fluor 488-phalloidin-labelled actin filaments of U2OS cells. From the captured raw hologram, the 3D spatial organizations and spatial-averaged molecular dipole orientations of the actin filaments within the sample volume are visualized simultaneously in the $xy$ plane max Intensity projected image (top row). The phase distribution of the hologram was numerical propagated, CAO optimized and then converted to obtain the dipole orientations distributions (Materials and methods). Binary masks that generated from the computational refocused intensity reconstructions were applied to the dipole orientations distributions to eliminate the interfering from random phase noise in the background (Materials and methods). Scale bar: 6.36-μm in the full field of view reconstructions (top row) and 1.77-μm in the region of interests (ROIs) 1&2.

The pSIDH 3D intensity reconstructions of the sample fluorescence clearly indicate that the spatial organizations of actin filaments in cell is 3D. By propagating the complex-valued hologram to a specific imaging depth $z_r$ within the sample volume, details on the spatial organizations of actin filaments which could be blurred and unresolvable because of the diffraction of light along $z$, were computationally brought back in pSIDH without scanning. It should be emphasized that in conventional microscopy this can be only accomplished with a relative low temporal resolution by capturing a mass of 2D images when scanning the sample or the objective lens along $z$. Meanwhile, the calibrated and converted phase distributions of the propagated fields provide insight into the spatial-averaged molecular dipole orientations on the refocused actin filaments, at all the possible imaging depth within the sample volume. In all the two selected ROIs in Fig. 6, the majority of the molecules were measured to orient along the actin filament axis, which is consistent with the results obtained in other polarization microscopic techniques[6,14]. Again, in conventional 2D polarization microscopy additional exposures are required to extract the desired dipole orientations at each z scanning position, which further decrease the overall imaging speed of the system.

**Discussion**

In summary, we have presented pSIDH to realize scanning-free holographic 3D polarization imaging of spatial incoherent illuminated or self-luminous samples. Theoretical analysis and experiments demonstrate a linear dependence of the phase values of the numerically refocused holographic reconstruction on the input polarization angle. Those enable the retrieval of polarization orientation with an statistical accuracy of about 0.5°. Meanwhile, the detected orientations are proved to be independent with the spatial location of the object, thus enables the polarization imaging of 3D object. All the results indicate that in pSIDH both the 3D coordinates of the sample and the polarization information of the light can be encoded in the captured hologram, and then decoupled from each other in the computational reconstructed complex field.

The application of pSIDH in fluorescence microscopy permits to evaluate the spatial organization of molecular dipole orientation of fluorescent labelled architectures in the cell. Our method has overcome the limitations of current scanning-based 3D polarization microscopy and opens a rapid and simple way where the structures and dipole orientations of the sample can be simultaneously reconstructed in 3D from only a single frame of complex-valued hologram. In our high-magnification, high NA pSIDH microscope one of the key points to guarantee the sufficient SNR of hologram, especially when imaging the fluorescence biological samples, is the using of high-efficiency planar liquid crystal holographic elements. It is important to note that SIDH visualization of biological sample have been performed using GRIN elements and custom-designed birefringent lens, without the discussion on polarization imaging[17,18,22] (though it is possible). The majority of PLL-based SIDH experiments, on the other hand, are performed under low NA so far[23–25]. Moreover, in our pSIDH microscope, image degradations that caused by ineluctable optical aberrations (e. g., spherical aberrations when oil immersion objective lens are used) can be addressed by CAO numerical 3D image optimization. These enables rapid, high-fidelity, scanning-free volumetric imaging of fluorescence sample with a spatial resolution of ~0.35-μm × 0.35-μm × 2.1-μm over an FOV of 113-μm × 113-μm × 100-μm, provides a 3D visualization of the spatial organization of actin filaments inside the U2OS cells.

The potential of our pSIDH microscope in measuring the spatial structures and ensemble molecular organization of protein assemblies is highlighted by imaging U2OS with their actin filaments labelled with Alexa Fluor 488-phalloidin. The spatial organizations of the actin filaments are obtained from the amplitude of the holographic reconstructed complex field, while the calibrated phase of the field provides a direct visualization on their spatial averaged molecular dipole orientations. For most of the molecules in Fig. 6, their spatial-averaged dipole orientations are measured to be paralleled with the actin filaments axis, which is consistent with the fact that the phalloidin molecules are oriented along the actin filament axis. However, diversities on the dipole orientations distribution were observed at some spatial locations on the actin filaments, which could come from the under estimating on the retrieved orientations that caused by the non-negligible degree of orientational flexibility of phalloidin molecules during the imaging integration time. We discussed theoretically the potential solution on this problem (Supplementary Notes). Another interesting source of this diversities on dipole orientations correlates with the potential specifical functional characteristics of the actin-microtubule crosstalk [26,27], which is a one of important future directions of our research.

CAO optimization provides significant improvements simultaneously on the spatial resolution, SNR of 3D reconstructed intensity distributions and the polarization orientation detection accuracy of pSIDH. Therefore, our computational image reconstruction and optimization framework shed light on a new adaptive optics method for multidimensional image optimization in polarization microscopy. Although further theoretical and experimental analysis on this field is out of the scope of this manuscript, recent develop methods [28] show that there is room for optimization in this direction. It should be emphasized that a key requirement for the utilization of 3D image reconstruction and CAO optimization is that the wavefront recording ability of the optical system. Among all the potential optical system and method (e.g., light-field microscopy), our pSIDH has shown advantages in the balancing between imaging speed, 3D spatial resolution, SNR, FOV and the complexity of computational algorithms, with the breaking superiority in concurrently measuring of the polarization orientations. About the potential under estimating on the orientations in pSIDH when considering the orientational flexibility of molecules, our theoretical analysis shows that, for the dynamic molecules their mean orientation angles $\rho$ and wobbling cone angle $\delta$ can be simultaneously measured by sequential capturing two holograms with different optical orientations of the PLL (as detailed in Supplementary Notes). The experimental demonstration on these is one of our future directions. Other alternatives such as multidimensional parameter estimating and image optimization based on deep learning also provides promising ways to push forward the performances our pSIDH microscopy in practical biological imaging.

Overall, pSIDH is a simple and high efficiency approach for wide-field, high fidelity scanning-free polarization volumetric fluorescence computational microscopy. We believe the rapid captured 3D polarization imaging datasets by pSIDH will be useful in not only uncovering the biophysical mechanisms behind essential biological processes such as the actin filaments driving mechanics in the cell development and immune response, but also the optical morphology and testing in the field of nanofabrication and material science.

**Materials and methods**

**pSIDH performance validation system.** Measurements are carried out on a proof-of-principle pSIDH system to demonstrate its validation and basic performances in scanning-free 3D polarization orientation imaging. Briefly, referring to the optical setup in Fig. **S1**, the sample is illuminated by the spatial incoherent source consisting of a LED source (528-nm central wavelength, GCI-060403, DHC) and a diffuser. The transmitted light is then collimated by an objective lens (f = 100-mm, GCL-010205, DHC), passes through the tube lens (f = 50.8-mm, GCL-010204, DHC) and planar liquid crystal lens (PLL, PBL25-532-F100, LBTEK) combination, and provides the self-interference patterns on the camera (4.65-µm pixel size, DCU224M, Thorlabs). A rotational linear polarizer is used before the camera to generate the phase-shifting holograms necessary to suppress the DC term and twin images in the 3D reconstructions. In the dual-channel 3D polarization imaging experiments, an additional sample and illumination path that uses identical type of LED illumination was introduced before the objective lens and combined using a beam splitter.

**pSIDH microscope.** A pSIDH microscope was designed based on a homebuilt fluorescence wide-field microscope. A diode-pumped solid-state laser (457-nm central wavelength, 85-BLT-605, Melles Griot) was used as excitation light source. The laser was coupled into a single-mode fiber (GCX-XSM-4/125-FC/PC-FC/PC, DHC) and subsequently collimated (by an f = 40-mm achromatic lens, Linos) again after the fiber output to get a uniform illumination with diameter of 10-mm. The laser was cleaned up by a linear polarizer (GCL-050003, DHC) and polarized by a quarter wave plate (AQWP05M-580, Thorlabs) oriented such as to obtain a quasi-isotropic excitation in the sample plane. To overfill the effective field of view of the objective, the beam was further expanded by a 4f telescope system (two achromatic lenses with focal length of f = 125-mm and f = 500-mm respectively, AC254-125-A and ACT508-500-A, Thorlabs). After a excitation filter (MF469-35, Thorlabs), a dichroic mirror (MD498, Thorlabs) reflects the laser toward the home-build inverted microscope body (combination of a 45° mirror, an objective lens holder (CSA1100, Thorlabs) and a vertical stage (M-MVN80, Newport)), followed by a tube lens (f = 200-mm, AC254-200A, Thorlabs) to focus the beam in the back focal plane of the oil immersion objective lens (UPlanFLN 100X, NA=1.3, Olympus), to provide a wide-field epi-illumination. A single-axis piezo-z-scanning stage (P-736 ZR2S, PI) was used for capturing of 3D wide-field image stacks. The fluorescent from the sample was collected by the same objective, separated from the excitation laser by the dichroic mirror, and then filtered by a band pass emission filter (MF525-39, Thorlabs). In the fluorescence imaging path, the back pupil of the objective lens was imaged onto an achromatic doublet lens (f = 200-mm, GCL-010617A, DHC) and PLL (PBL25-532-F2500-SP, LBTEK) combination that provides the differentially focused beams necessary to achieve fluorescence self-interference. Another important function of PLL is the encoding of polarization orientation in addition to the 3D coordinates of the sample. For a polarized light, after passing through the PLL, two beams that have different wavefront curvatures and orthogonal polarization state (specifically, RCP and LCP, as detailed in [23–25]) are obtained. The differences in wavefront curvatures are correlated with the 3D coordinates of the sample, while the phase shift between the RCP and LCP components are direct depends on the polarization orientation of the light (as detail in Supplementary Note). The optical conjugation of the back pupil plane to the location of the PLL is of important to produce an optimal overlapping of the two beams that produce the self-interference patterns. The wide-filed images and pSIDH holograms were captured by a EMCCD camera (13-μm pixel size, iXon Ultra 888, Andor). The overall magnification of the holographic imaging system is 117×, corresponding to a pixel size of 110.7-nm in sample space. The distance between the camera and the PLL-doublet combination is chosen according to the optimal holographic recording condition (as detailed in [16]. In our system, distances between optical elements: 320-mm from tube lens to PLL-doublet combination, 95-mm from doublet to PLL inside the combination, resulting in a 170-mm optimal recording distance in our system), although minor degradations on the imaging performance is observed when this optimal condition can not be satisfied for all the possible imaging depth for a 3D sample. For the scanning-free 3D pSIDH dipole orientation imaging of phalloidin-labelled U2OS cells, a combination of two doublet lenses (f = 200-mm and f = 400-mm respectively, GCL-010617A and GCL-010610A, DHC) was used together with the PLL. This provides a smaller spacing factor *s* [22] which benefit the holographic recording of the fine structures that have size far smaller than the diffraction-limit (e.g., 5 ~ 9-nm for a single F-actin filament [29]).

**Image acquisition, hardware control and data processing.** To generate the interference patterns, the orthogonal RCP and LCP polarized beams from LPP are projected to the same linear polarization direction by the rotational linear polarizer before the camera. Additionally, during the holographic image acquisition four different polarizer orientation angles of 0°, 45°, 90° and 135° are used to introduce the desired phase-shifting values of 0, π/2, π and 3π/2 between those two beams. The orientation angle of the polarizer is controlled by a custom-designed motorized rotation lens mount. The rotation of polarizer is controlled by the Arduino board and synchronized with image exposure of the camera. The hardware control is accomplished in μManager. The complex-valued hologram $H$ is obtained by combing the captured phase-shifted holograms $I_1$, $I_2$, $I_3$, and $I_4$ as $H = (I_1-I_3) - j(I_2-I_4)$. To calibrate the image drifts, the phase-shifted holograms were registered using Fiji plugin of Register Virtual Stack Slices[30]. 3D complex sample fields are then reconstructed by stacking the 2D reconstructions that obtained by numerical back-propagating the complex-valued hologram to different axial distances. The axial magnification of the SIDH visibly exhibits a nonlinear dependence on the imaging depths, therefore leading to an anisotropic voxel depth in the 3D holographic reconstructions when the reconstruction distances with fixed steps are used. These is calibrated by using a depth-dependent-nonlinear-spaced reconstruction distances that obtained by manually aligning the intensity of the holographic 3D reconstructed field with the referenced wide-field image stack of the same sample. The final voxel size of holographic 3D reconstruction is 110.7-nm × 110.7-nm × 400-nm (73.8-nm × 73.8-nm × 267-nm for the U2OS cells experiments). To determine the accurate reconstruction distance of the 2D sample, auto-focusing algorithm in where image sharpness and statistical analysis on the gradients of image gray level were used as metric. Calibration is required to convert from the phase of the holographic reconstructed field to the polarization orientation angles of the sample light. In the proof-of-principle experiments (Fig. 1) these is implemented by recording several holograms under different input orientations when an object with well-defined and controllable orientation angles is used in the system. For the results in Fig. 1, the phase value $\varphi$ (rad) of the holographic reconstructed field are converted into the polarization orientation angle $\rho$ (degree) of the input light via $\rho = (-\varphi/2 + 0.03727)/ 0.01695$, which is obtained by a linear fitting on the data shown in Fig. 1a. The pSIDH microscope is calibrated in a similar way to extract the mean dipole orientations averaged during the imaging integration time, where $\rho$ = (-$\varphi$/2 + 0.05024)/ 0.0175 (Supplementary Fig. S**5**). For the data of phalloidin-labelled U2OS cells, background noise was subtracted from the 3D holographic intensity reconstructions. An additional linear scaling factor and offset was introduced into the phase calibration because of the complexity in the polarization state of the emitted fluorescence. The linear scaling factor and offset were kept the same for all the 3D spatial locations in the reconstruction volume. Binarizations were implemented on the refocused 2D intensity reconstructions to extract the spatial areas that correspond to the in-focused actin filaments. The resulting binary masks were applied to the retrieved dipole orientation distributions to separate the signal from the random phase noise in the background. The background subtraction, binarization and masking were accomplished in Fiji [30]. The back-propagating and phase calibration algorithms were implemented through customized MATLAB (Mathworks) scripts.

**Computational adaptive optics (CAO) aberration correction of the 3D holographic reconstruction.**
To compensate the image degradation caused by system- and/or sample-introduced optical aberrations, an image optimization on the calibrated 3D complex-valued holographic reconstructions is implemented by the CAO algorithm as detailed in our previous work [19,20]. Briefly, a Zernike-mode compensation phase mask is first generated and then applied to the holographic reconstructions in the Fourier domain, and an iterative image optimization procedure is implemented in where the specific Zernike coefficients that provides an optimized image quality of the compensated images are obtained. For a simple task where only one or a few kinds of Zernike aberrations need to be considered (e.g., to correct only the spherical aberrations as shown in Fig. 5), a manual CAO optimization is used. For a more complex sample and imaging condition, the first 20$^{th}$ Zernike polynomials are optimized in parallel through SPGD-CAO as detailed in [19] to provide a better imaging performance. The optimization can be applied either to the entire FOVs for isotropic aberration correction, or separately to different ROIs (as shown in Fig. 5a, b) when the optical aberrations are anisotropic especially in 3D imaging of biological samples. For 3D image stack the above optimizations are individually applied to each slice. The CAO algorithms were implemented through customized MATLAB (Mathworks) scripts.

**Sample preparation. 3D fluorescent particle samples.** For the alignment and performance validation of wide-field and pSIDH microscope, 3D sample was prepared by mixing 500-nm diameter fluorescent beads (1:200 v/v, 505/515, F8813, Invitrogen), and agarose (1:40 w/v) with water.

**U2OS cells.** 24 mm-diameter high-precision (170-μm) glass coverslips (0117640, Marienfeld) were cleaned with base piranha (Milli-Q water, 30% ammonium hydroxide, 35% hydrogen peroxide at a 5:1:1 volume ratio) for 15 min, rinsed with Milli-Q water for 2 × 5 min in a bath sonicator, sonicated in 70% ethanol for 5 minutes, and air-dried before coating with fibronectin (F1141, SIGMA) for 2 h at room temperature (RT) and at a final fibronectin concentration of 20 μg/mL in PBS. U2OS cells were seeded onto fibronectin-coated coverslips and allowed to spread for overnight on. Cells were fixed with 4% paraformaldehyde-PBS, rinsed for three times with PBS, permeabilized with 0.1% Triton X-100/PBS, rinsed for three times with PBS, blocked in 3% goat serum in 0.1% (v/v) Triton X-100/PBS, rinsed for three times with PBS, and stained with the indicated primary antibodies overnight at 4 °C followed by incubation with fluorescent conjugated secondary antibodies. and with three 10-min washes in-between antibody incubations. and then mounted with DAPI-containing mounting medium (P0131, Beyotime).

**Availability of data and materials**

The datasets used and/or analysed during the current study are available from the corresponding author on reasonable request.

**Funding**

National Natural Science Foundation of China (61575009), Natural Science Foundation of Beijing Municipality (4182016), Beijing Municipal Natural Science Foundation (3222001), Applied Basic Research Fund of the School of Physics and Optoelectronic Engineering, Beijing University of Technology (ABRFSPOE05), Young Elite Scientist Sponsorship Program by BAST (BYESS2023066), Tianjin Key Laboratory of Micro-scale Optical Information Science and Technology.

**Authors' contributions**

T. M. conceived the research. T. M. built the optical system and control software. T. M., W. Z. and L. Z. conducted the experiments and analyzed the results. T. M., R. Z., X. L., H. H., H. Z. and Y. W. prepared the manuscript. R. Z., X. L., and H. H. prepared the biological samples. All the authors contributed to the manuscript. T. M. and Y. W. supervised the research.

**Acknowledgements**

We acknowledge the helps of Prof. Z. Zhang for his constructive suggestions. The Y. Wan group acknowledges the support of NSFC. T. Man acknowledges the support of BAST.

**Competing interests**

The authors declare that they have no competing interests.